\documentclass[10pt, conference]{IEEEtran}
\usepackage{blindtext, graphicx}
\usepackage{amsmath}
\usepackage{algorithm}
\usepackage[noend]{algpseudocode}
\usepackage{cite}
\usepackage[colorlinks=true,citecolor=blue,pdftex]{hyperref}

\let\tempone\itemize
\let\temptwo\enditemize
\renewenvironment{itemize}{\tempone\addtolength{\itemsep}{0.5\baselineskip}}{\temptwo}

\begin{document}

\title{DCRoute: Speeding up Inter-Datacenter Traffic Allocation while Guaranteeing Deadlines}

\author{
\IEEEauthorblockN{Mohammad Noormohammadpour}
\IEEEauthorblockA{% Ming Hsieh Department of Electrical Engineering\\
University of Southern California\\
noormoha@usc.edu}
\and
\IEEEauthorblockN{Cauligi S. Raghavendra}
\IEEEauthorblockA{% Ming Hsieh Department of Electrical Engineering\\
University of Southern California\\
raghu@usc.edu}
\and
\IEEEauthorblockN{Sriram Rao}
\IEEEauthorblockA{%Cloud and Information Services Lab\\
Microsoft\\
sriramra@microsoft.com}
}

\maketitle

\begin{abstract}
Datacenters provide the infrastructure for cloud computing services used by millions of users everyday. Many such services are distributed over multiple datacenters at geographically distant locations possibly in different continents. These datacenters are then connected through high speed WAN links over private or public networks. To perform data backups or data synchronization operations, many transfers take place over these networks that have to be completed before a deadline in order to provide necessary service guarantees to end users. Upon arrival of a transfer request, we would like the system to be able to decide whether such a request can be guaranteed successful delivery. If yes, it should provide us with transmission schedule in the shortest time possible. In addition, we would like to avoid packet reordering at the destination as it affects TCP performance. Previous work in this area either cannot guarantee that admitted transfers actually finish before the specified deadlines or use techniques that can result in packet reordering. In this paper, we propose DCRoute, a fast and efficient routing and traffic allocation technique that guarantees transfer completion before deadlines for admitted requests. It assigns each transfer a single path to avoid packet reordering. Through simulations, we show that DCRoute is at least $200$ times faster than other traffic allocation techniques based on linear programming (LP) while admitting almost the same amount of traffic to the system.
\end{abstract}

\begin{IEEEkeywords}
Datacenter; Routing; Traffic Allocation; Traffic Scheduling; Deadlines; Wide Area Networks;
\end{IEEEkeywords}

\section{Introduction}
Cloud Computing allows customers to build online applications that can cost effectively scale as necessary \cite{cc}. It provides a massive pool of resources for online applications that can be flexibly obtained when needed and then returned back to the pool at a later time. Such resources can be provided to different applications on top of the infrastructure built and maintained by a cloud company. Examples of hosted online applications include video streaming, data storage and sharing, and big data processing. Most companies that provide cloud computing services own multiple datacenters placed in different cities and countries in order to improve availability, reduce end-to-end delays for end users, and provide customized regional services. At the time of writing this paper, Amazon has more than two dozen availability zones each consisting of one or more discrete datacenters \cite{aws}, Microsoft has 22 regions with plans to build 8 more \cite{azure}, and Google relies on more than a dozen datacenters \cite{google}. 

Many applications hosted on these datacenters need to transfer data to their peers in other datacenters for the purpose of data replication and synchronization \cite{sync, amoeba}. The aim is to improve fault-tolerance and user quality of service by making multiple copies of data and getting data closer to end users. Most of these transfers have to be completed before a deadline in order to meet customer service level agreements (SLAs) and they can take hours to complete \cite{tempus}. For example, search engines may have to exchange data among datacenters in order to synchronize their search databases and storage applications may need to back up user data over certain periods.

As mentioned in \cite{swan}, inter-datacenter traffic can be categorized into three groups: \textit{interactive} traffic that has to be transmitted as soon as possible since it is in the critical path of user experience, \textit{elastic} traffic that requires timely delivery which can be modeled in the form of a deadline, and \textit{background} traffic which is bandwidth hungry and has less priority than the other two categories of traffic. In this paper, we focus on elastic traffic with user specified deadlines. 

Previous work in the context of inter-datacenter traffic scheduling either fails to consider the negative effects of packet reordering caused by multiplexing packets over different paths (AMOEBA \cite{amoeba}) or cannot guarantee that admitted requests will actually complete transmission before the deadlines specified by customers (B4 \cite{b4}, SWAN \cite{swan}, TEMPUS \cite{tempus}). In addition, AMOEBA and TEMPUS which are the state-of-the-art techniques in this area, model the allocation problem as large linear programs (LP), with possibly hundreds of thousands of variables, solving which incurs large memory and CPU overhead and can take a long time.

Avoiding packet reordering allows data to be instantly delivered to applications upon arrival of packets. In addition, inter-datacenter networks have characteristics similar to WAN networks (including asymmetric link delays and large delays for links that connect distant locations) for which multiplexing packets over different paths has been shown to considerably degrade TCP performance \cite{wan-reordering}. Putting out of order packets and segments back in order can be expensive in terms of memory and CPU usage, especially when transmitting at high rates.

As explained in \cite{real-mptcp}, TCP needs to buffer as much as the bandwidth-delay product ($BDP$) of the network in lossless and $2 \times BDP$ in lossy networks to put out of order packets back in order. For high speed inter-datacenter networks with tens of gigabits of speed and tens of milliseconds of latency considerable amount of buffering may be needed. In \cite{juggler}, authors show how Vanilla kernel uses $50\%$ more CPU in presence of severe packet reordering and present Juggler, a reordering resilient network stack designed for low latency datacenter networks which can reduce the extra CPU utilization to $10\%$: still a considerable amount. For higher latency networks, due to large variation of RTT over multiple paths, reordering may further increase CPU utilization.

In \cite{rcd}, we proposed RCD, a technique that speeds up the traffic allocation problem by scheduling transfers close to their deadlines. Through simulations, we showed that RCD speeds up the allocation process by allowing new transfers to be scheduled only considering the residual bandwidth which would result in creation of much smaller LP models. However, we did not discuss the reordering problem and only evaluated our technique for a single link scenario.

In this paper, we propose DCRoute, a fast and efficient routing algorithm which eliminates the need for LP modeling and: 

\vspace{0.5em}

\begin{itemize}
    \item \textbf{Guarantees that admitted transfers complete prior to specified deadlines.}
    \item \textbf{Schedules all packets of each transfer over the same path to avoid packet reordering.}
    \item \textbf{Works much faster than other techniques while admitting almost equal traffic to the system.}
\end{itemize}

\vspace{0.5em}

The input to DCRoute is the network topology as well as a list of transfers including their volumes and deadlines which are submitted to DCRoute in the order of arrival. DCRoute assigns a single path to every transfer and generates a transmission schedule which specifies the rate at which every transfer should be sent over their assigned path during current timeslot. In this paper, we verify the performance of DCRoute through long running simulations. In practice, label switching (such as VLAN tagging) can be used to enforce paths and rate-limiting at the end hosts can be used to enforce transmission rates as in SWAN \cite{swan}.

\section{Problem Description}
\begin{figure}
    \centering
    \includegraphics[width=0.7\columnwidth]{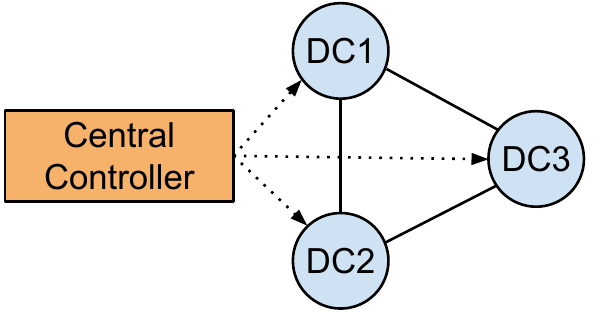}
    \caption{Problem Setup}
    \label{problem-setup}
\end{figure}

Figure \ref{problem-setup} shows our problem setup which is comprised of multiple datacenters in different locations managed by a central controller. The datacenters are connected using high speed WAN links. Applications hosted on these datacenters will make transfers that may take hours to complete and have to be finished before specified deadlines. Due to large transfer time, the time it takes for the source datacenter to request a path from the central controller and the time it takes to setup such a path is considered negligible. However, the time it takes for the scheduling algorithm to prepare a traffic schedule depends on the scheduling algorithm which we aim at minimizing. In addition, in order to avoid packet reordering, we would prefer to send all packets of a transfer on a single path. 

As in \cite{amoeba, swan}, we assume the timeline is divided into properly sized timeslots over which the transmission rate is constant. Using a slotted timeline allows for schedules with variable transmission rates over time. In our model, an inter-datacenter request $R$ is represented with four parameters $src(R)$, $dst(R)$, $vol(R)$ and $dl(R)$ which stand for source, destination, volume of data to be transferred and the deadline prior to which the transfer has to be completed.

Similar to previous work \cite{amoeba}, we assume arriving requests are put into a queue and processed in the order of arrival and that there is no preemption: once a request is allocated, it cannot be unallocated from the system. At each moment, we have two parameters $t_{now}$ and $t_{end}$ which represent current timeslot and the latest deadline among all active requests, respectively. A request arriving sometime in timeslot $t$ can be allocated starting timeslot $t+1$ since the schedule and transmission rate for current timeslot is already decided upon and broadcast into all datacenters. Also, at any moment $t$, $t_{now}$ is the timeslot that includes $t$ (current timeslot), and $t_{now}+1$ is the next available timeslot for allocation (next timeslot). 

Upon arrival of a request, a central controller decides whether it is possible to allocate it considering some criteria that includes the total available bandwidth over future timeslots. If there is not enough room to allocate a request, the request is rejected and can be submitted to the system again later with a new deadline. 

A request is considered \textbf{active} if it is accepted into the system and its deadline has not passed yet. Some active requests may take many timeslots to complete transmission. The total unsatisfied demand of an active request is called the residual demand of that request. 

\begin{figure}
    \centering
    \includegraphics[width=0.9\columnwidth]{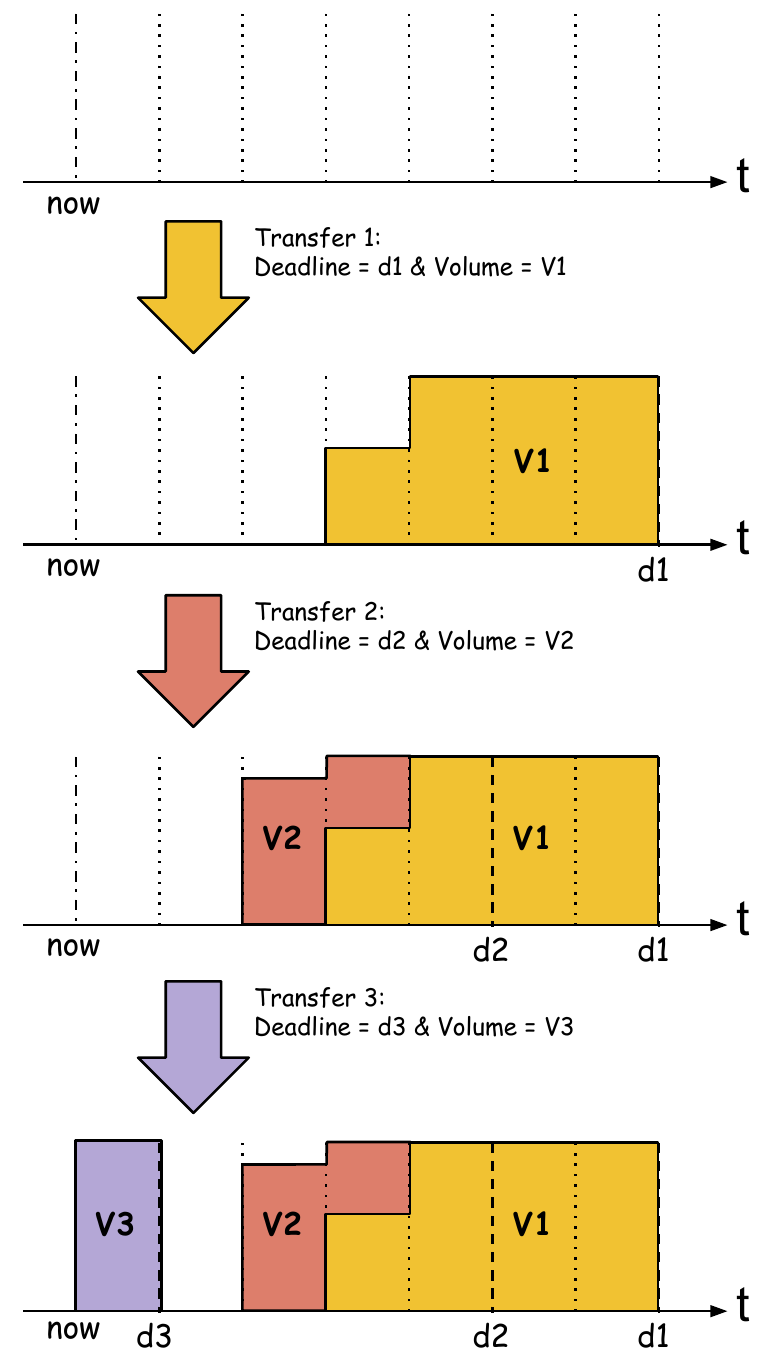}
    \caption{An example of ``As Late As Possible" allocation}
    \label{ALAP}
\end{figure}

\textbf{Allocation Problem:} \textit{Given active requests $R_1$ through $R_n$ with residual demands $D_1$ to $D_n$ ($0 \le D_i, ~ 1 \le i \le n$), is it possible to allocate a new request $R_{n+1}$? If yes, what is a possible schedule?}

An important characteristic of network traffic is that the size of smallest traffic unit (which is a packet) is significantly smaller than link capacities (which are in the range of gigabits nowadays). This allows us to solve the allocation problem by forming a linear program (LP) considering capacity constraints of the network edges as well as demand constraints of requests. The answer will give us a possible allocation if the constructed LP is feasible. Although this solution maybe straightforward, considering the number of active requests, number of links in network graph, and how far we are planning ahead into the future ($t_{end}$), the resulting LP could be large and may take a long time to solve. 
One of the ways to speed up this process is to limit the number of possible paths between every pair of nodes \cite{tempus}, for example, using k-shortest paths \cite{amoeba}. While solving the LP, another speedup method is to limit the number of considered active requests based on some criterion \cite{amoeba} such as having a common link with the new request. It is also possible to use customized iterative methods to solve the resulting LP models faster based on the solutions of previous LP models in a way similar to the water filling process \cite{tempus}. 

\emph{Our proposition is to avoid building an LP model in the first place by trying to allocate new requests only knowing the residual bandwidth on the edges for different timeslots.}

\section{DCRoute}
DCRoute relies on the following three techniques:

\vspace{0.5em}

\begin{itemize}
    \item \textbf{Requests are initially allocated as late as possible (ALAP) \cite{alap}} 
    \item \textbf{Utilization is maximized by pulling traffic from closest future timeslots into the upcoming timeslot} 
    \item \textbf{A variant of BFS search is used for path selection}
\end{itemize}

\vspace{0.5em}

Figure \ref{ALAP} provides an example of the ALAP allocation technique. As can be seen, when the first transfer is received the timeline is empty and therefore it is allocated adjacent to its deadline. The second transfer is allocated as close as possible to its deadline. The benefit of this type of scheduling is that requests do not use resources until it is absolutely necessary. This means resources will be available to other requests that currently demand them. Now when the third transfer is submitted, the resources are free and it just grabs as much bandwidth as needed. If we had allocated the first two requests closer to current time we may have had to either reject the third transfer or move the first two transfers ahead freeing resources for the third transfer.

Assume requests $R_1$ through $R_n$ are current active requests and we would like to allocate $R_{n+1}$. For every request $R_i,~ 1 \le i \le n$, we either have $dl(R_i) \le dl(R_{n+1})$ or $dl(R_i) > dl(R_{n+1})$. In the former case, since there is no preemption, there is no way to increase the chance of new request being accepted by shifting the traffic allocation of request $i$ away as it has to be completed before $dl(R_{n+1})$. For the latter case, since we allocated all requests ALAP, the traffic is already shifted out of $R_{n+1}$'s window as much as possible. As a result, it is possible to decide on admission of new request by just looking at the residual bandwidth on the links. For a single link, since all requests use a single shared resource (link capacity) this technique allows us to optimally decide whether a new request can be allocated. For a network, each request is routed on multiple links and there are many ways to schedule requests ALAP. If some link is used by multiple requests that are routed on different edges, how traffic is allocated on the common link can affect multiple other links which will affect the requests that use those links later on. Despite this uncertainty, we will show that using this feature, we can greatly speed up the allocation process.

Using ALAP alone can result in poor utilization as we always push traffic towards future timeslots and leave the current timeslot underutilized. In order to maximize utilization, upon beginning of a new timeslot, the scheduler looks at future timeslots and pulls as much traffic as possible from the closest timeslots in the future to the upcoming timeslot. Pulling from closest timeslot allows the ALAP characteristic of allocation to hold true afterwards: all residual demands will still be allocated as close to their deadlines as possible.

While pulling traffic from future timeslots, a request can span over multiple links and if there is some traffic fully occupying the next timeslot on one of those links, it is impossible to pull traffic back for that request from the future timeslots to the next timeslot. In such cases, we may need to pull traffic from requests that may not be the closest to current timeslot. This can result in an allocation in which some requests are not scheduled ALAP because they can be pushed further into the future. To address this problem, after pulling traffic from future timeslots, we sweep requests allocated on future timeslots and push them forward as much as possible. 

\begin{figure*}
    \centering
    \includegraphics[width=\textwidth]{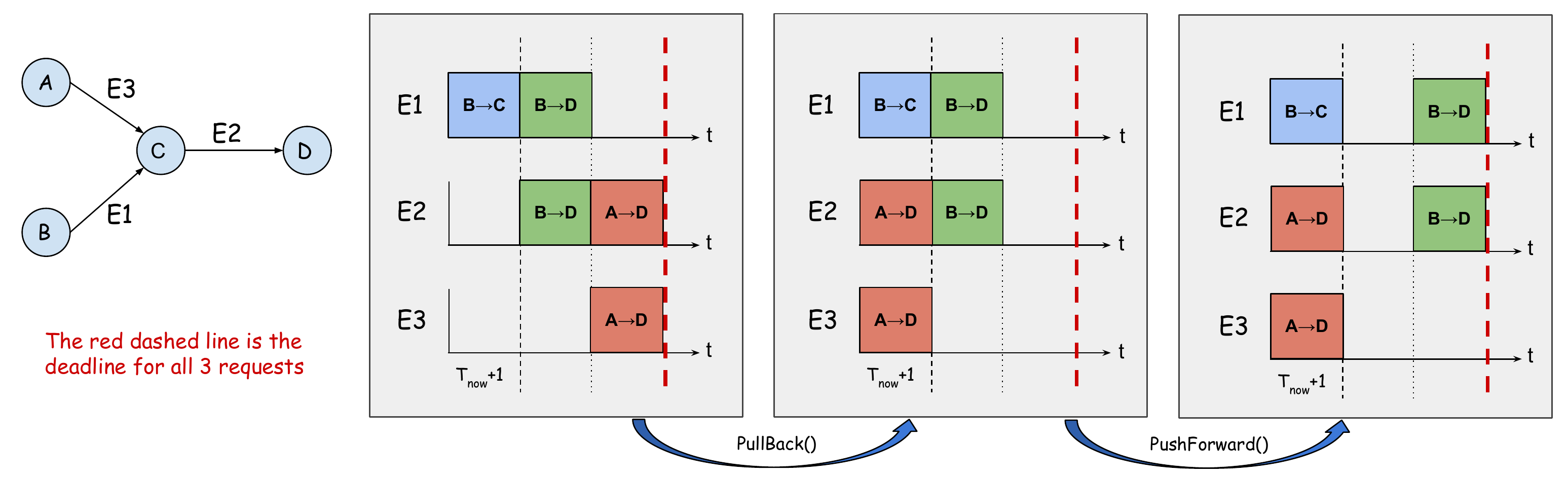}
    \caption{An example of improving utilization while keeping the final allocation ALAP}
    \label{ex}
\end{figure*}

Figure \ref{ex} shows an example of this process. There are three different requests all of which having the same deadline. It is not possible to pull back the green request as the link $E1$ is already occupied. Therefore, we have to pull the orange request (PullBack phase). Afterwards, the allocation is not ALAP anymore, so we push the green request towards its deadline (PushForward phase). The final allocation is ALAP and the utilization of upcoming timeslot is maximum.

\subsection{DCRoute Algorithms} 
We assume a graph $G(V,E)$ connecting datacenters with $M$ bidirectional links. For simplicity, we also assume all links/edges have equal capacity of $1.0$. Every edge $e$ has a boolean \textit{use} property which identifies whether that edge can be used in the course of routing. Also each node $v$ has $4$ parameters \textit{v}, \textit{r}, \textit{rv} and \textit{rs}. If $p$ is the path on BFS tree from src($R$) ending at $v$, these parameters represent whether $v$ has been visited before, the number of hops to $v$ from source, the load of bottleneck link, and the sum of loads on all edges from source to $v$, respectively. Moreover, we have variables $S_{t,m}$ that represent the total sum of traffic over link $m$ from time $t_{now}+1$ to $t$. Every time a new request is submitted to the system, $t_{end}$ is updated so that it covers all active requests. Finally, we define the \textbf{active window} as the set of all timeslots over all edges from time $t_{now}+1$ to $t_{end}$. Our algorithms only operate on the active window.

\textbf{Allocate($R$):} Algorithm \ref{algo1} is executed upon arrival of a new request $R$ and performs multiple BFS searches. In every search, we calculate multiple costs, remove edges with highest costs, and compare the result with previous steps. While examining different paths, our heuristic finds the path with most preferred characteristics: the total sum of traffic before dl($R$) over the chosen path will be minimal compared to other paths when $R$ gets allocated on that path. 

In each round, the path assignment algorithm starts by choosing the path with the least number of hops and calculates the total cost of sending the request on that path. Also, the bottleneck link on that path is identified. If the total cost is less than the best path found in previous steps, this path replaces the best path. Next, all edges with an equal or higher cost than the bottleneck edge are removed from the network. This process continues until there is no path from source to destination. Now, if it is possible to allocate the request on the selected path, we apply the allocation, otherwise, the request is rejected since admitting it might result in many more future requests to be rejected.

\begin{algorithm}
    \scriptsize
    \caption{}
    \label{algo1}
    \begin{algorithmic}[1]
    \Procedure{Allocate($R$)}{}
        
        \For {$t=t_{end} + 1$ \textbf{to} $R.dl$}
            \For {$e \in E$}
                \State $S_{t,e} \gets S_{t-1,e}$
            \EndFor
        \EndFor
        
        \State $t_{end} \gets$ \textbf{max}($\emph{R.dl}, t_{end}$)
        
        \For {$e \in E$}
            \State $e.use \gets True$
        \EndFor
        
        \For {$v \in V$}
            \State \{$v.v$, $v.r$, $v.rv$, $v.rs$\} $\gets$ \{$False$, $0$, $0$, $0$\}
        \EndFor
        
        \State $SE \gets$ array of edges sorted by $S_{R.dl,e}$ descending
        
        \State $i \gets 0$, $j \gets 1$, $flag \gets True$, $r_{b} \gets$ BFS\_test($R$)

        \While{$r_{b}.r$ \textbf{is not} $-1$}
            \If {$(r_{b}.rv > 0)$ \textbf{and} $flag$}
                \State $flag \gets False$
                \While {$S_{R.dl,SE[i]} > r_{b}.rv$}
                    \State $SE[i].use \gets False$
                    \State $i \gets i+1$
                \EndWhile
            \EndIf

            \State $SE[i].use \gets False$, $i \gets i+1$

            \State $r_{n} \gets$ BFS\_test($R$)
            \If {$r_{n}.r == -1$} 
                \textbf{break}
            \EndIf

            \State $\alpha \gets r_{b}.r \times R.vol + r_{b}.rs$
            \State $\beta \gets r_{n}.r \times R.vol + r_{n}.rs$

            \If {$(\alpha > \beta)$ \textbf{or} $((\alpha == \beta)$ \textbf{and} $(r_{b}.rv > r_{n}.rv))$} 
                \State $j \gets i+1$, $r_{b} \gets r_{n}$, $flag \gets True$
            \EndIf
        \EndWhile
        
        \While {$i \ge j$}
            \State $i \gets i-1$, $SE[i].use \gets True$
        \EndWhile
        
        \State $path \gets$ \textit{edges of the path ending at $R.dst$ on the BFS}
        \State \hspace{3.5em} \textit{tree starting at $R.src$} 
        \If{PathAllocate($path,R,False$)}
            \State \textbf{return} PathAllocate($path,R,True$)
        \Else 
            \State \textbf{return} $False$
        \EndIf
    \EndProcedure
    \vspace{1.5em}
    \Procedure{BFS\_test($R$)}{}
        \For{$v \in V$}
            \State \{$v.v$, $v.r$, $v.rv$, $v.rs$\} $\gets$ \{$False$, $0$, $0$, $0$\}
        \EndFor
        \State $Q \gets$ a queue having $R.src$ as first element
        \State $R.src.v \gets True$
        \While{$Q$ is not empty}
            \State $node \gets$ head of the $Q$ removed
            \If {$node == R.dst$}
                \State \textbf{return} $\{node.r, node.rv, node.rs\}$
            \EndIf
            \For{$next \in$ all neighbors of $node$}
                \State $e \gets$ edge connecting $node$ to $next$
                \If{$e.use == True$ \textbf{and} $next.v == False$}
                    \State add $next$ to $Q$
                    \State $next.v \gets True$ 
                    \State $next.r \gets next.r$ $+$ $1$
                    \State $next.rv \gets$ \textbf{max}$(node.rv, S_{R.dl, e})$
                    \State $next.rs \gets node.rv+S_{R.dl, e}$
                \EndIf
            \EndFor
        \EndWhile
        \State \textbf{return} $\{-1, -1, -1\}$
    \EndProcedure
    \vspace{1.5em}
    \Procedure{PathAllocate($path,R,apply$)}{}
        \State $vol \gets R.vol$
        \For {$t=R.dl$ \textbf{to} $t_{now}+2$ \textbf{step} $-1$}
            \State $space \gets vol$
            \For {$e \in path$}
                \State $space \gets$ \textbf{min}$(space,1-S_{t_{now}+1,e})$
            \EndFor
            \If {$space > 0$} 
                \State $vol \gets vol-space$
                \If{$apply$}
                    \For {$e \in path$}
                        \State add $space$ of $R$ to edge $e$ at time $t$
                        \For{$t'=t$ \textbf{to} $t_{end}$}
                            \State $S_{t',e} \gets S_{t',e}+space$
                        \EndFor
                    \EndFor
                \EndIf
            \EndIf
        \EndFor
        \State \textbf{return} $vol == 0$
    \EndProcedure
    \end{algorithmic}
\end{algorithm}

\textbf{PullBack():} Algorithm \ref{algo2:1} looks at the timeslots starting $t_{now}+2$ to $t_{end}$ and pulls back traffic to $t_{now}+1$ (next timeslot to be scheduled). When pulling back traffic, all edges on a request's path have to be checked for unused capacity and updated together as we pull traffic back.

\begin{algorithm}
    \scriptsize
    \caption{}
    \label{algo2:1}
    \begin{algorithmic}[1]
    \Procedure{PullBack()}{}
        \For {$t=t_{now}+1$ \textbf{to} $t=t_{end}$}
            \For {$e \in E$}
                \State $reqs \gets$ all requests allocated on $e$ at $t$
                \For{$R \in reqs$}
                    \State $vol \gets$ how much of $R$ allocated on $e$ at $t$
                    \State $path \gets$ path assigned to $R$ upon allocation
                    \For{$e' \in path$}
                        \State $vol \gets$ \textbf{min}$(vol,1-S_{t_{now}+1,e'})$
                    \EndFor
                    \If{$vol > 0$} 
                        \For{$e' \in path$}
                            \State move $vol$ of $R$ from edge $e'$ at $t$
                            \State to edge $e'$ at $t_{now}+1$
                            \For{$t' = t_{now}+1$ \textbf{to} $t' = t$}
                                \State $S_{t',e'} \gets S_{t',e'}+vol$
                            \EndFor
                        \EndFor
                    \EndIf
                \EndFor
            \EndFor
        \EndFor
    \EndProcedure
    \end{algorithmic}
\end{algorithm}

\textbf{PushForward():} After pulling some traffic back, it may be possible for some other traffic to be pushed ahead even further. Algorithm \ref{algo3:1} scans all future timeslots starting $t_{now}+2$ and makes sure that all demands are allocated ALAP. If not, it moves as much traffic as possible to the future timeslots until all residual demands are ALAP.

\begin{algorithm}
    \scriptsize
    \caption{}
    \label{algo3:1}
    \begin{algorithmic}[1]
    \Procedure{PushForward()}{}
        \For {$t=t_{now}+2$ \textbf{to} $t=t_{end}$}
            \For {$e \in E$}
                \State $reqs \gets$ all requests allocated on $e$ at $t$
                \For{$R \in reqs$}
                    \State $vol \gets$ how much of $R$ allocated on $e$ at $t$
                    \State $path \gets$ path assigned to $R$ upon allocation
                    \For {$t_2=R.dl$ \textbf{to} $t_2=t+1$ \textbf{step} $-1$}
                        \State $free \gets$ \textbf{min}$(vol,$ free space on $e$ at $t_2)$
                        \For{$e' \in path$}
                            \State $free \gets$ \textbf{min}$(free,$ free space on $e'$ at $t_2)$
                        \EndFor
                        \If{$free > 0$} 
                            \For{$e' \in path$}
                                \State remove $free$ of $R$ from edge $e'$ at $t$
                                \State add $free$ of $R$ to edge $e'$ at $t_2$
                                \For{$t' = t$ \textbf{to} $t' = t_2$}
                                    \State $S_{t',e'} \gets S_{t',e'}-free$
                                \EndFor
                            \EndFor
                        \EndIf
                    \EndFor
                \EndFor
            \EndFor
        \EndFor
    \EndProcedure
    \end{algorithmic}
\end{algorithm}

\textbf{Walk():} Algorithm \ref{algo4:1} is executed when the allocation for next timeslot is final. This algorithm tells each datacenter of the decided allocation and adjusts request demands accordingly by deducting what is scheduled to be sent from the total demand.

\begin{algorithm}
    \scriptsize
    \caption{}
    \label{algo4:1}
    \begin{algorithmic}[1]
    \Procedure{Walk()}{}
        \State \textbf{Broadcast the schedule for $t_{now}+1$ to all datacenters}
        \For{$e \in E$}
            \State $D \gets S_{t_{now}+1}$
            \For{$t = t_{now}+1$ \textbf{to} $t = t_{end}$}
                \State $S_{t, e} \gets S_{t, e} - D$
            \EndFor
        \EndFor
        \State $t_{now} \gets t_{now}+1$
        \State $t_{end} \gets$ \textbf{max}$(t_{end}, t_{now}+1)$
    \EndProcedure
    \end{algorithmic}
\end{algorithm}

\subsection{DCRoute and Multi-Path Routing}
As mentioned earlier, to avoid packet reordering, DCRoute maps every transfer to exactly one path and if there is no single path that can allocate a transfer, it will be rejected. Although this sounds too restrictive, we show in the next section that limiting every transfer to a single path results in $2\%$ less utilization in the worst case.

To use the aggregate bandwidth on multiple paths, one can use Multi-Path TCP (MPTCP) \cite{mptcp} which allows sending traffic from multiple interfaces of a single host. MPTCP is based on multiple sub-flows that behave similar to single TCP flows. While using MPTCP, the overall CPU and memory footprint can increase significantly compared to TCP which is the cost paid for increased bandwidth. However, by carefully choosing which parts of the data goes over what path, it is possible to improve the CPU footprint \cite{real-mptcp}.

To increase network utilization, it is possible to use MPTCP and apply DCRoute to all sub-flows created. To do that, one can divide the total demand of a request over multiple sub-flows and assign the same deadline as the deadline of the original request to all of them. If all sub-flows can be allocated with DCRoute, then the request is accepted.

Deciding on the number of sub-flows and the portion of traffic that goes over each one of them is not a trivial problem. It may be possible to extend DCRoute by ranking the paths found in \textbf{Allocate} procedure and choosing a subset of the best paths. Further discussion of this subject is out of the scope of this paper. 

\section{Simulation Results}
In this section, we perform simulations to evaluate the performance of DCRoute. We generate synthetic traffic requests with Poisson arrival and input the traffic to both DCRoute and a few other techniques that can be used for traffic allocation. Two metrics are being measured and compared: \textbf{allocation time} and \textbf{portion of rejected traffic} both of which are desired to be small.

\textbf{Simulation Parameters:} We used the same traffic distributions as described in \cite{amoeba}. Requests arrive with Poisson distribution of rate $\lambda$. Also, total demand of each request $R$ is distributed exponentially with mean $\frac{1}{8}$ proportional to the maximum transmission volume possible prior to $dl(R)$. In addition, the length of requests is exponentially distributed for which we assumed a mean of $10$ timeslots. We performed the simulations over $500$ timeslots.

All simulations were performed on a machine with an Intel Core i$7$-$6700$T CPU and $24$ GBs of memory and the algorithms are coded in Java. To solve linear programs faster, we used Gurobi Optimizer \cite{gurobi} with a free academic license. Gurobi Optimizer can speedup the LP solving process using several techniques including parallel processing. All simulations presented here are performed $3$ times and the average is reported. We compare DCRoute with the following allocation schemes for all of which we used the same objective function as \cite{amoeba}:

\textbf{Global LP:} This technique is the most general and flexible way of allocation which routes traffic over all possible edges. All active requests are considered for all timeslots on all edges creating a potentially large linear program. The solution here gives us a lower bound on traffic rejection rate.

\textbf{K-Shortest Paths:} Same as Global LP, however, only the K-Shortest Paths between each pair of nodes are considered in routing. The traffic is allocated using a linear program over such paths. We simulated four cases of $K \in \{1, 3, 5, 7\}$. It is obvious that as $K$ increases, the overall rejection rate will decrease as we have higher flexibility for choosing paths and multiplexing traffic.

\textbf{Pseudo-Integer Programming (PIP):} In terms of traffic rejection rate, comparing DCRoute with the previous two techniques is not fair as they allow multiplexing packets on multiple paths. \textbf{The aim of this technique is to find a lower bound on traffic rejection rate when all packets of each request are sent over a single path.} To do so, the general way is to create an integer program involving a list of possible paths (maybe all paths) for the new request and fixed paths for requests already allocated. The resulting model would be a non-linear integer program which cannot be solved using standard optimization libraries available. We instead created a number of linear programs each assigning one of the possible K-Shortest Paths for the newly arriving request. We then compare the objective values manually and choose the best possible path. In our implementation, we chose $K = 20$. This $K$ seems to be more than necessary as we saw negligible improvement in traffic rejection rate even when increasing $K$ from $5$ to $7$. Using PIP, the path over which a request is transferred is decided upon admission and does not change afterwards. We implemented two versions of this scheme:

\begin{itemize}
    \item \textbf{Pure Minimum Cost (PMC):} We choose the path that results in smallest objective value.
    \item \textbf{Shortest Path, Minimum Cost (SPMC):} Amongst all shortest paths that result in a feasible solution and have the least number of hops, we choose the one with smallest objective value.
\end{itemize}

\subsection{Google's GScale Network}
\begin{figure*}
    \centering
    \includegraphics[width=\textwidth]{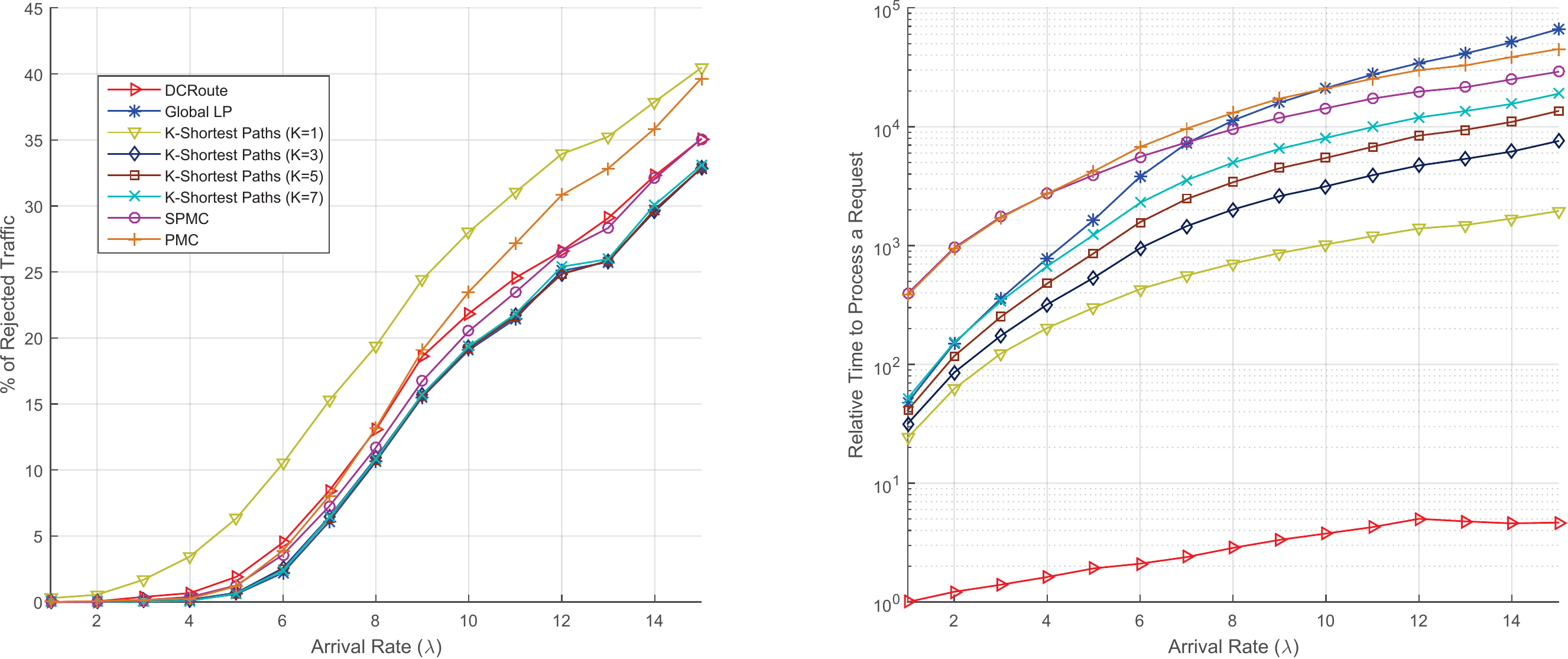}
    \caption{Total \% of rejected traffic and relative request processing time for GScale network with 12 nodes and 19 links}
    \label{115}
\end{figure*}

GScale network \cite{b4} comprised of $12$ nodes and $19$ links (at $2013$, they have $15$ datacenters as of $2016$ \cite{google}) is a private network that connects Google data centers. We used the same topology to evaluate DCRoute as well as other allocation schemes. Figure \ref{115} shows the rejection rate of different techniques for different arrival rates from low load ($\lambda=1$) to high load ($\lambda=15$). We have included the schemes that potentially multiplex traffic over multiple paths just to provide a lower bound. Comparing with PMC and SPMC schemes over all arrival rates, DCRoute performs $< 2\%$ worse than the one with minimum rejection rate. Also, compared to all schemes, DCRoute rejects at most $4\%$ more traffic. 

Figure \ref{115} shows the relative time to process a request using different schemes. This time is calculated dividing the total time to allocate/adjust all requests over all timeslots by the total number of requests. DCRoute is about $3$ orders of magnitude faster than either PMC or SPMC. It should be noted that the rate at which time complexity grows drops as we move towards higher arrival rates since there is less capacity available for new requests and many arriving requests get rejected by failing simple capacity constraint checks.

\subsection{Variable Network Size}
We simulated different methods against four networks from $5$ to $20$ nodes: 

$(N, M) \in \{(5, 7), (10, 17), (15, 27), (20, 37)\}$

In our topology, each node was connected to $3$ or $4$ other nodes at most $2$ hops away. The arrival rate was kept constant at $\lambda=6.0$ for all cases.

\begin{figure*}
    \centering
    \includegraphics[width=\textwidth]{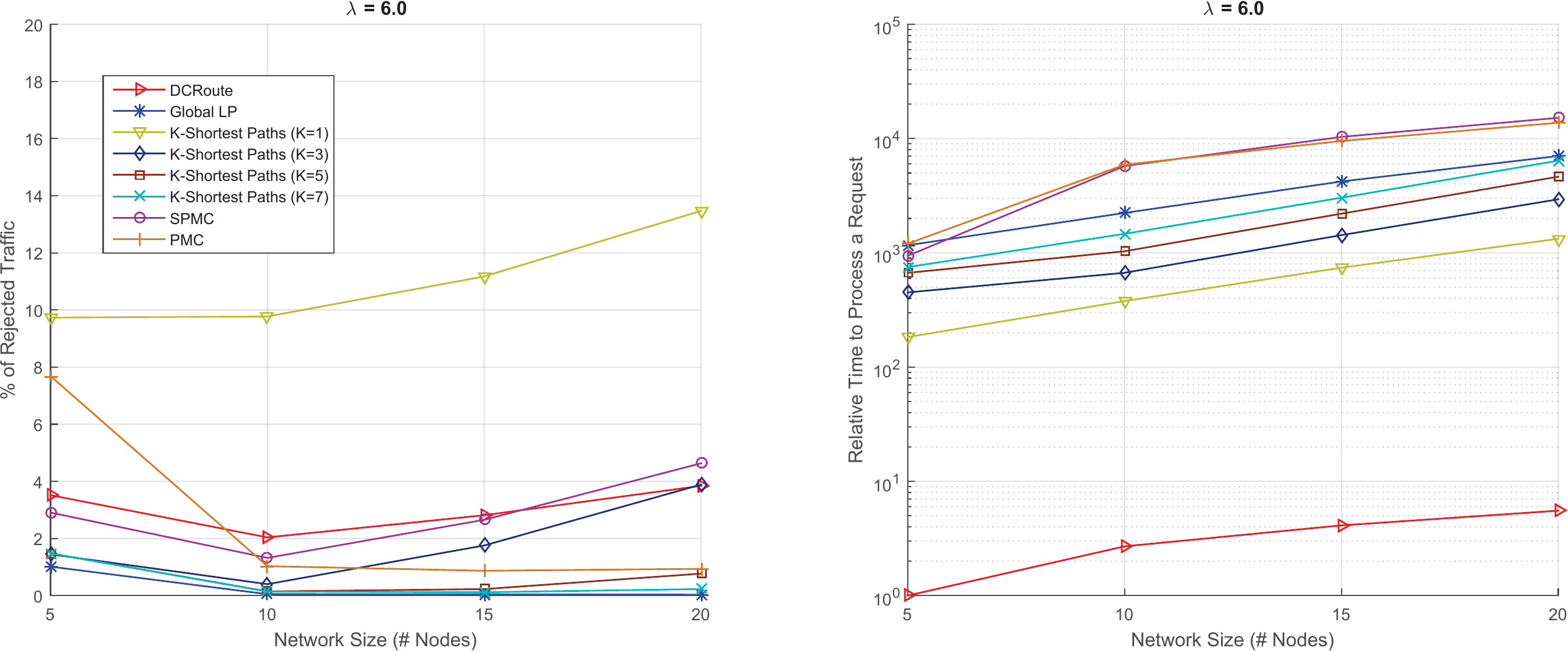}
    \caption{Total \% of rejected traffic and relative request processing time for different network sizes}
    \label{520}
\end{figure*}

Figure \ref{520} shows the rejection rate of different schemes for different network sizes. As network size increases, since $\lambda$ is kept constant, the total cpacity of network increases compared to the total demand of requests. As a result, for a scheme that multiplexes request traffic over different paths, we expect to see decrease in rejection rate. For the K-Shortest Paths case with $K \in \{1,3\}$ we see increase in rejection rate which we think is because these schemes cannot multiplex packets that much. Increasing the network size for these cases can cause more requests to have common links as the network is sparsely connected and create more bottlenecks resulting in a higher rejection rate. 

PMC has a high rejection rate for small networks since choosing the minimum cost path might result in selecting longer (more hops) paths that create larger number of bottlenecks due to collision with other requests. Increasing network size, there are more paths to choose from and that results in less bottlenecks and therefore less rejection rate. In contrast, SPMC enforces the selection of paths with smaller number of hops resulting in lower rejection rates for small networks (due to request paths colliding less) and more rejections as network grows due to less diversity of chosen paths.

Compared to these two approaches, DCRoute balances the choice between smaller and longer paths. The assigned path has the least sum of load on the entire path and the least bottleneck load among all such paths. Paths with heavily loaded links and unnecessarily larger number of hops are avoided. As a result, rejection rate compared to \textit{min(PMC, SPMC)} is relatively small ($<3\%$) for all network sizes. Also, as Figure \ref{520} shows, similar to previous simulation, DCRoute is almost three orders of magnitude faster than PIP schemes and more than $200$ times faster than all considered schemes.

\subsection{Effect of Timeslot Length}
As mentioned earlier, the timeline is divided into timeslots. In this paper, we do not discuss the exact duration of timeslots, however we would like to see how smaller timeslots affect the amount of rejected traffic as well as the speed of different methods. The transmission rate of transfers only changes when moving from one timeslot to the next. Having timeslots that last longer than most requests would essentially result in a fixed transmission rate for such transfers. This causes less critical transfers (with a later deadline) to have to share the bandwidth with more critical transfers (with a closer deadline), reducing the flexibility of the system and increasing the number of rejected transfers. Having very short timeslots on the other hand can result in unnecessary scheduling overhead and practical complications. For example, rate-limiters at the end hosts may not be able to converge to specified rate if they have to change rate too often. Therefore, a proper timeslot length could be a few times smaller than most requests but large enough to avoid unnecessary processing overhead.

\begin{figure*}
    \centering
    \includegraphics[width=0.99\textwidth]{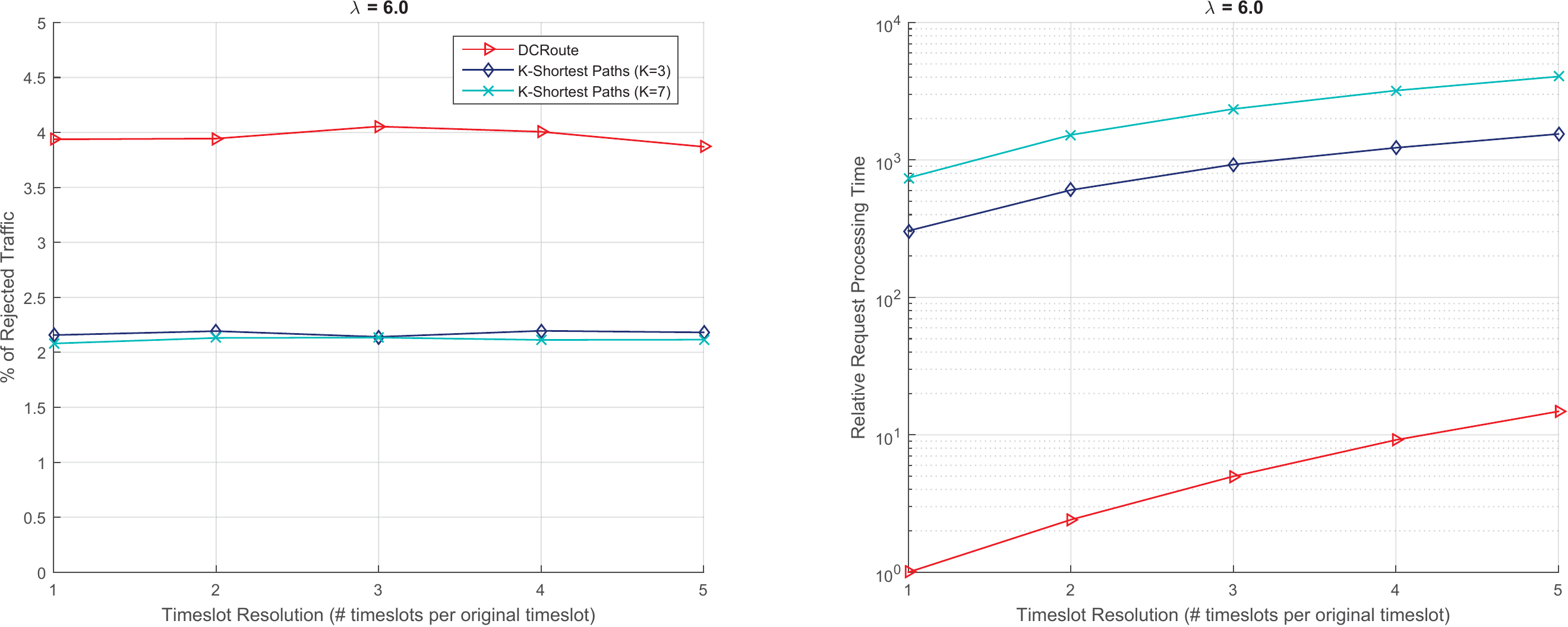}
    \caption{Total \% of rejected traffic and relative request processing time as timeslot resolution increases}
    \label{51}
\end{figure*}

As shown in Figure \ref{51}, we divide every timeslot into $2$ to $5$ smaller timeslots. We again use GScale topology and consider $(\lambda = 6)$. We only considered the K-Shortest Paths techniques with $K \in \{3, 7\}$ since during previous tests they provided the best utilization at the least processing cost. It appears that if the timeslots are short enough, making them shorter does not improve the load accepted into the system: DCRoute rejects $2\%$ more traffic compared to K-Shortest Paths schemes for all timeslot resolutions. Although DCRoute is always more than $2$ orders of magnitude faster than K-Shortest Paths schemes, it does have a slightly higher growth rate in processing time as the timeslot resolution increases.

\subsection{Discussion}
We do not directly compare our scheme with any of the previous schemes, such as AMOEBA or TEMPUS, since we pursue the following two objectives together which is not the case for previous work:

\begin{itemize}
    \item Avoiding any packet reordering
    \item Guaranteeing deadlines for admitted transfers
\end{itemize}

However, since AMOEBA is the most similar to DCRoute, we would like to provide an approximate comparison. We argue that AMOEBA is essentially an extension to the K-Shortest Paths LP technique $(K = 10)$ introduced here. By intelligently avoiding LP modeling for some special cases, AMOEBA provides a speedup of $30$ times compared to 10-Shortest Paths LP for $(\lambda = 8)$ \cite{amoeba}. For the same arrival rate and request volume/deadline distribution, DCRoute performs more than $1000$ times faster than 10-Shortest Paths LP technique. Also in a previous work called RCD \cite{rcd}, we showed how the close to deadline scheduling technique can speed up the traffic allocation by up to $15$ times compared to AMOEBA while resulting in same utilization over a single link. DCRoute is based on RCD coupled with a path selection heuristic that eliminates the need for LP modeling.

\section{Conclusions and Future Work}
In this paper, we proposed DCRoute, a routing algorithm for Inter-Datacenter networks which guarantees that transfers complete before their deadlines and to avoid reordering, schedules all packets of a request on the same path. Inspired by the fact that allocating ALAP allows for new requests to be scheduled considering only residual bandwidth, DCRoute performs much faster than schemes based on LP modeling. It is more than $2$ orders of magnitude faster than all simulated schemes and $3$ orders of magnitude faster than schemes that do not multiplex transfer packets on multiple paths.

We showed that DCRoute admits at most $4\%$ less traffic compared to schemes that multiplex packets on multiple paths (which provide an estimate of lower bound on rejected traffic) and $2\%$ less traffic compared to schemes that schedule all packets of each transfer on the same path.

Finally, we studied the effect of timeslot resolution and found out that making timeslots smaller than necessary does not increase the admitted load and only incurs extra processing costs. In this paper, we evaluated DCRoute with synthetic traffic. Evaluation of DCRoute using real inter-datacenter traffic is suggested as a future work. In addition, studying the effects of link failures and developing methods to properly handle them can be a subject for future work.

\section*{Acknowledgement}
We would like to thank the anonymous reviewers of HiPC whose suggestions and comments helped us improve the quality of this paper.

{\footnotesize \bibliographystyle{unsrt}
\bibliography{citations.bib}}

\end{document}